\documentstyle[aps,epsfig,multicol]{revtex}

\newcommand\rt{\rightarrow}

\newcommand{\be}{\begin{equation}}
\newcommand{\ee}{\end{equation}}
\newcommand{\bea}{\begin{eqnarray}}
\newcommand{\eea}{\end{eqnarray}}

\newcommand{\brho}{\mbox{\boldmath $\rho$}}

\input{epsf.tex}

\def\figone#1#2#3{\begin{figure}[!ht]
\centering \leavevmode
\epsfxsize=0.9\columnwidth \epsfbox{#1}
\caption{\small #2 \label{#3}}
\end{figure}
}

\begin{document}

\title{Topological jamming of extended structures and the glass transition}
\author{Dibyendu Das, Jane$^{'}$ Kondev and Bulbul Chakraborty}
\address{Martin Fisher School of Physics, Brandeis University, 
Mailstop 057, Waltham, Massachusetts 02454-9110, USA}
\maketitle

\begin{abstract}
We propose a new scenario for glassy dynamics in frustrated systems with no 
quenched-in randomness, based on jamming of extended dynamical structures 
near a critical point. This route to a glassy state is demonstrated in a 
lattice model of fluctuating lines. 
Numerical simulations of the model show non-exponential relaxations and 
diverging energy barriers in the vicinity of a thermodynamic phase transition. 
A master equation for the coarse grained dynamics is constructed. It   
shows how topological jamming leads to the observed glassy dynamics.

\vskip0.5cm
%\noindent PACS numbers: {}

\end{abstract}

\begin{multicols}{2}

Foams, granular materials and supercooled
liquids, exhibit a transition from a flowing liquid phase to a jammed
phase\cite{Edwards_Grinev}.  The dynamics of the flowing state near
the transition are characterized by slow relaxations mediated by
large-scale spatial rearrangements. These  {\em dynamical
heterogeneities} were  dramatically demonstrated
in recent experiments on colloids near the glass transition\cite{Weitz}. 
An important question posed by these observations  
is: What is the relation between length and time scales in a glass 
forming material?

The mode-coupling theory predicts a divergence of relaxation 
time scales with no dramatic growth of length scales. This is born out by experiments. 
A scaling theory of the glass transition, based on the
existence of a zero-temperature critical point\cite{Sethna}, leads to a
similar conclusion about the relation between length and time scales. 
The frustration-limited domain theory\cite{Kivelson} relates  the origin
of the dynamical heterogeneities and the activated nature of the
dynamics to a critical point ``avoided'' due to frustrations imposed by 
competing interactions.  Similarly, the random first-order transition 
theory\cite{wolynes} 
relies on the presence of frustration in glass formers and suggest 
that an unusual critical point is responsible for the dynamical behavior 
observed near the glass transition. 

A phenomenological jamming phase diagram was recently proposed as a 
means of putting under one umbrella the whole spectrum of systems exhibiting dynamical 
arrest\cite{Nagel}. How, and if the above mentioned  theories relate to jamming 
ideas remains to be explored. 
Here we propose a scenario for glassiness based on the interplay of  
jamming and an underlying critical point. We  demonstrate it in detail in  a 
lattice model of fluctuating lines.

\paragraph{Loop model:}

In constructing a lattice model aimed at describing supercooled
liquids, we concentrated on two key features; frustration and the
presence of extended structures.  A model which has these attributes
is the three coloring representation of the loop model\cite{kondev1}.
The model is defined on a $2$-dimensional hexagonal lattice, in which 
all bonds are colored $A$, $B$ or $C$ with the
constraint that no two bonds which share a vertex are of the same
color.  The number of colorings is exponential in the number of
vertices and the system has an extensive entropy. This entropy is
maximized when the number of bonds of any color, in each of the three
directions is the same. This entropy was calculated exactly by
Baxter\cite{baxter}.  A coloring with all the bonds in one direction
having one and the same color has zero entropy density. By introducing
an energy parameter that favors colorings in the zero entropy sector a
phase transition can be induced in the model.

The frustration of the coloring model gives rise to naturally
occurring extended structures in the system. Consider all the $A$ and
$B$ colored bonds: there is one of each at every vertex, and therefore
they must form loops. 
All the properties of the model, including the correlations of the
microscopic variables, the colors, can therefore be traced back to the
behavior of the ensemble of loops.  These loops do not intersect, they
have orientation ($ABAB\ldots$ versus $BABA\ldots$), and therefore
they can be thought of as contour lines of a height field, that lives
on the dual lattice\cite{kondev1} and provides an appropriate order
parameter for the loop model.

The lattice with periodic boundary conditions, has a linear dimension
$L$ with $2 L^2$ sites and $3 L^2$ bonds and the three lattice
directions are indexed by $\alpha = 1, 2, 3$.  We introduce a
long-range interaction energy between $C$ colored bonds, with a
parameter $\mu$ measuring the strength of this interaction:
\be
E = - {\mu \over L^2}{\sum_{\alpha = 1,2,3} (2 N_{C}^{\alpha} - L^2)^2} \ .  
\label{energy}
\ee   
Here $N_{C}^{\alpha}$ is the total number of $C$ bonds in the $\alpha$
direction. In the ground state all of the $C$ colored bonds are aligned in one of 
the three directions. This ordering effect acts against the 
entropic influence which tries to keep the number of colors homogeneous in 
all directions, and hence disordered. 
As the value of the coupling constant $\mu$ is increased 
from $0$, the system undergoes a phase transition from a disordered 
to an ordered phase. 

The phase transition is best described in terms of the height
representation. In the disordered state, where the colors are randomly
arranged, the corresponding surface is {\it rough} and {\it untilted};
Fig. \ref{height}$(a)$ shows the height of a typical coloring
state. In the ordered state, with no extensive entropy, the surface is
{\it smooth} and maximally {\it tilted} as shown in Fig. 
\ref{height}$(b)$.
\figone{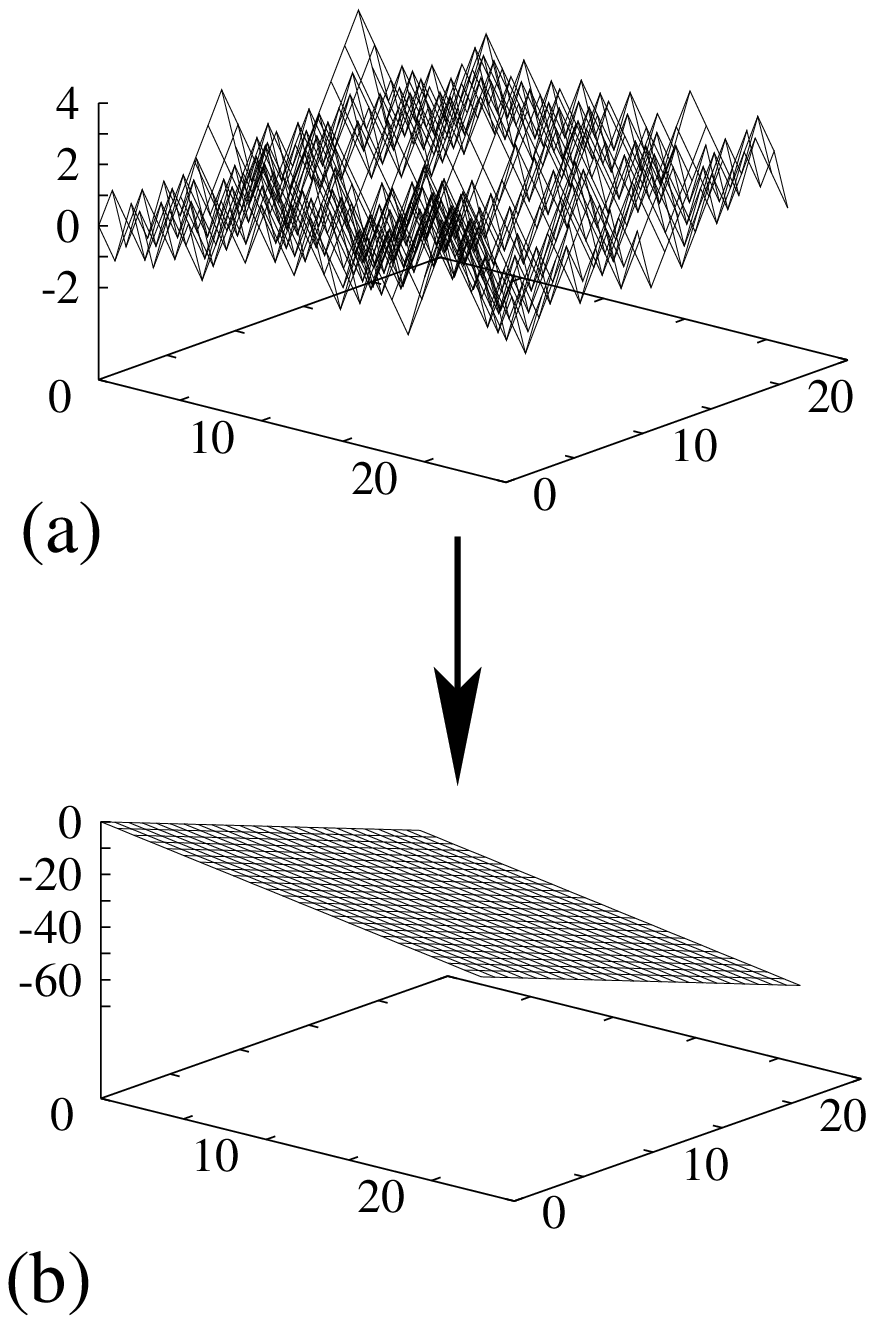}{The height profiles in the (a) disordered, and (b) 
ordered states are shown, for $L = 24$; the $x-$ and $y-$axes 
denote the lattice coordinates, and $z-$axis gives the height.}{height}

A tilting transition of the surface occurs at $\mu = \mu_*$. The
existence of a transition at a finite value, $\mu_*$ can be argued as
follows.  The entropy is maximum for a rough, untilted
surface\cite{baxter}. Similarly from Eq.~\ref{energy} it follows that 
the energy is also maximized at zero tilt. 
%and for small value of the tilt, $\brho$.  Here
%$\brho = (\rho_x,\rho_y)$, is the tilt vector. Focusing on simple
%tilts with only one non-vanishing component, $\rho$, the entropy can
%be expanded as $S(\rho) \simeq S(0) - b
%\rho^{2}$.
%From Eq.~\ref{energy} it follows that the energy is also maximized at
%zero tilt, and it has a quadratic form: $E(\rho) = E(0) - 8 \mu
%\rho^2$. 
Therefore the free energy, $F = E - S$, which is concave at
$\mu = 0$, must change its curvature at a finite $\mu_*$.  
We numerically measured the equilibrium
probability distribution of tilts for different values of $\mu$,
and observed that the distribution becomes broader with increasing
$\mu$. Quadratic fits indicate a $\mu_*
\simeq 0.35 \pm 0.05$, at which the second derivative of the free energy
function at zero-tilt vanishes. A tilting transition at a finite
$\mu_*$ has been explicitly demonstrated in the two-color version of
the current model\cite{huiyin} lending further support to our
argument.

\paragraph{Loop dynamics:}

The
tilting transition, whose dynamics we are interested in studying,
involves a change in the density of non-zero winding number loops:
structures which span the system.  
%A local dynamics based on updating
%the bond colors and respecting the constraint is unable to produce a
%system spanning loop and create a tilt.  The minimal fluctuations
%which can lead to a change in the tilt involve loops and we,
%therefore, define our dynamics on the loops (mesoscopic) rather than
%the bond colors (microscopic).  
The existence of these mesoscopic
structures and the role they play in the tilting transition is
reminiscent of the extended dynamical structures observed in atomistic
simulations
\cite{donati} and experiments\cite{Weitz} near the glass transition.  
This motivates us to consider loop dynamics.

In a unit microstep in our simulation, we randomly choose a site
and choose either an $A-B$, $B-C$ or a $C-A$ loop passing through it,
and we switch the colors on every alternate bond, using ordinary
metropolis rules. For example, a $...ACAC...$ loop becomes a
$...CACA...$ loop, with probability $1$ if the energy change is
non-positive and with rate $\exp(-\Delta E)$ when $\Delta E$ is
positive.
Since the energy function defined in Eq. \ref{energy} depends on
global numbers and not on local ones, it is easy to see that all local
loop updates keep $\Delta E = 0$. Only when system spanning loops
with nonzero winding numbers are chosen for update, does the
possibility of $\Delta E \neq 0$ arise. Therefore, 
the tilting of the surface and hence the lowering of energy can
only be brought about by updates of system-spanning loops. This path
downhill in energy is blocked by barriers which lead 
to spectacularly slow dynamics near $\mu_*$, as will be shown below.

\paragraph{Simulation results:}We measure the equilibrium, macroscopic 
tilt-tilt autocorrelation function, corresponding to the slowest mode in the
system:
\be
C(t) = {{\langle (\brho(t + t_o) - 
{\langle \brho \rangle}).(\brho(t_o) - {\langle \brho \rangle}) \rangle}
 \over {{\langle (\brho(t_o) - {\langle \brho \rangle})^{2}\rangle}}}
\label{corr}
\ee 
Here $\brho = (\rho_x,\rho_y)$ is the tilt vector, and $\langle \cdots 
\rangle$ denotes an average over various initial times $t_o$ in a history. 
We obtain an average $C(t)$ by averaging several ($\sim 10$) histories. 
The time 
is measured in units of Monte-Carlo step per site (MCS). In
Fig. \ref{autocorr} we show $C(t)$ at different $\mu$ for $L=24$.
Similar curves have been obtained for two other system sizes, $L=36$
and $L=48$. The autocorrelations functions exhibit a two step-decay,
which may be termed the $\beta -$ and $\alpha -$ relaxations following
the standard terminology of mode-coupling theory\cite{mode_coupling}.
Although we do not attempt a detailed analysis of  the shape of these
relaxation curves in this letter, several features are apparent from Fig.
\ref{autocorr}: with increasing $\mu$ the time scale of relaxation to the 
plateau, the plateau height, and the time scale of the decay of the plateau 
increase monotonically. 
In fact the decay time scale of the plateau $\tau$ 
(see inset of Fig. \ref{autocorr}), 
grows exponentially and diverges at a finite
$\mu$ with the  divergence being well described by the Vogel-Fulcher form;
$\exp(A(L)/(\mu_*(L) - \mu))$ for all three system sizes studied. The extrapolation of $\mu_*(L)$ to the infinite system indicates a divergence close to the 
transition point $\mu_*$ obtained from the equilibrium distribution.
\figone{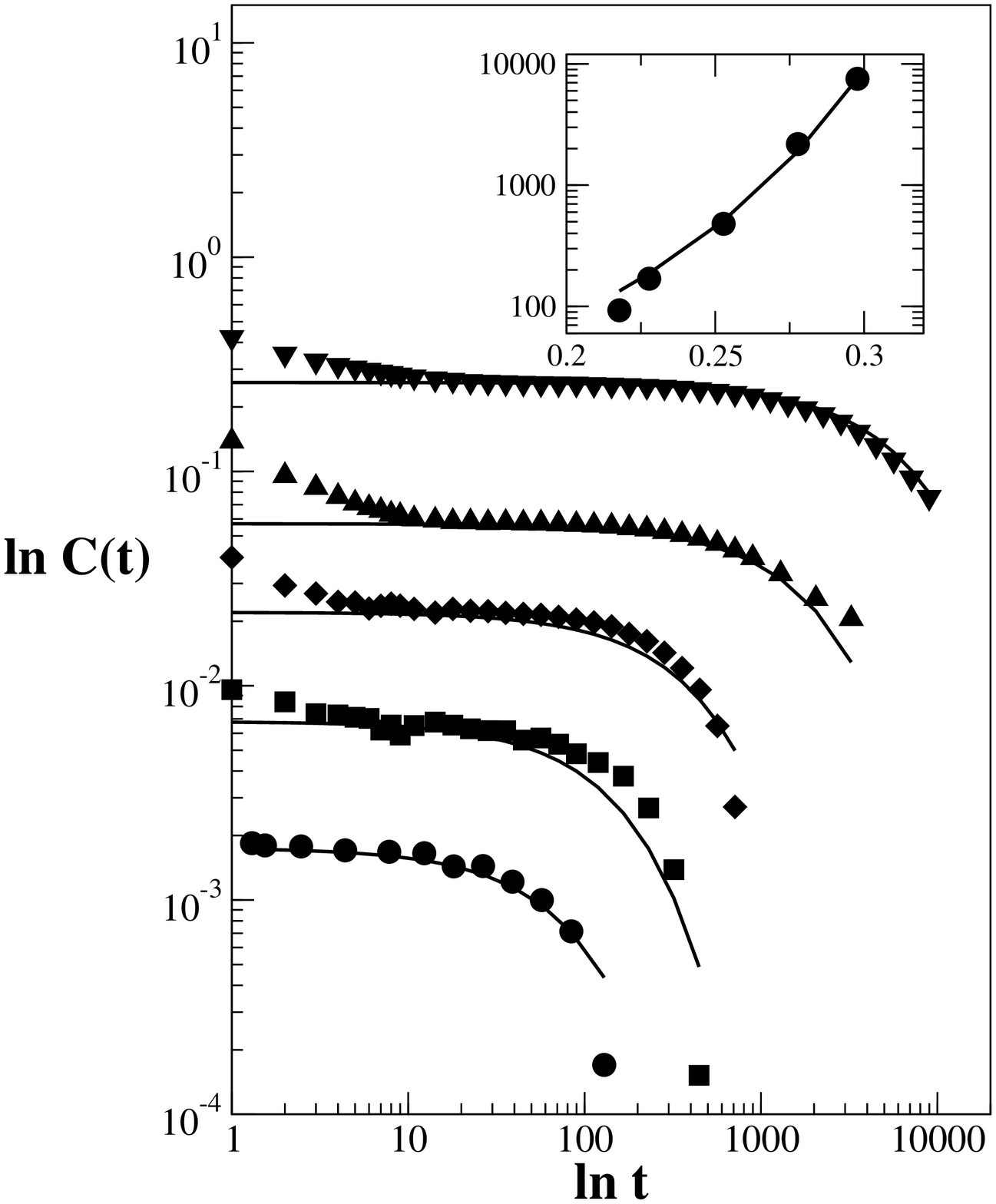}{The log-log plot of $C(t)$ versus $t$, at 
$\mu = 0.14$, $0.15$, $0.175$, $0.2$, and $0.22$ for $L = 24$. The
curves exhibit a 2-step decay, and the fits (shown as solid lines) provide us with  an
estimate of the decay time scale of the plateaus, $\tau$. The inset 
shows a plot of $\tau$ against $\mu$ and a 
Vogel-Fulcher fit to it.}{autocorr}

As discussed in detail in the following paragraphs, the exponential
divergence of $\tau$ and the shape of $C(t)$ can be associated with
jamming barriers and enhanced static fluctuations near the critical
point.

\paragraph{Residence time and transition rates of different tilted states:} 
The Vogel-Fulcher divergence of the relaxation time is a generic feature
of real glass formers. The origin of this anomalous slowing down
is still a matter of some debate.  
Within our simple model we investigate the crucial features
underlying the rapid rise in time scales, by  measuring 
the transition probabilities among different tilted states and  
constructing a coarse-grained dynamical model for the
tilt relaxation.  We find that the relaxation process is very
different from simple Langevin dynamics within a concave free
energy well and is, instead, dominated by traps and barriers.

Although the autocorrelation functions $C(t)$ were measured with
boundary conditions which allow the system to have any tilt $\brho$,  
%=\rho_x$ and $\rho_y$, 
to simplify our discussion, 
we concentrate on simple tilted states with the tilt pointing 
either in the $x$ or $y$ direction.
%direction such as $\rho_x =
%\rho$ and $\rho_y = 0$.
The measurements primarily involve studying
several histories (of the order of $10^4 - 10^5$) in which the system
starting from an initial state $\rho$, stays in that state for time
$t$ and makes a transition to a new state $\rho^{'}$ in the next time
interval $\delta t$. From these time histories, one can easily
calculate $P_{\rho \rt \rho}$, the probability that the system stayed
in the state $\rho$ for time $t + {\delta t}$, and
$P_{\rho \rt \rho^{'}}$, the probability that it stayed in the state
$\rho$ for time $t$ and made a transition to $\rho^{'}$ between $t$
and $t+{\delta t}$.

Assuming that the tilt configurations $\{\rho\}$ evolve via a Markov
process, the measured probabilities should have an exponential form:
$P_{\rho \rt \rho} = f_1 \exp(-t/{\tau_{\rho}})$ and $P_{\rho \rt
\rho^{'}} = f_2 \exp(-t/{\tau_{\rho}})$, where $f_1$ and $f_2$ are numbers less than unity.  From our extensive simulations we find that at each $\mu$
and for any $\rho$ (with $\rho \neq 0$) the probabilities $P_{\rho \rt
\rho}$ and $P_{\rho \rt \rho^{'}}$ (for say $\rho^{'} = (\rho + 1)$
and $(\rho - 1)$) have a common exponential factor with identical
decay constants $\tau_{\rho}$. This evidence validates the above
description in terms of the coarse-grained states $\{ \rho \}$.  The
$\rho = 0$ states are an exception since we found that $P_{0 \rt 0}$ 
is a stretched exponential in $t$. This observation has little bearing 
on our current discussion, and will be taken up in the future.     
%we will exclude them from 
%our subsequent discussion, and their effect on the dynamics will be 
%discussed in future work.

Based on the above arguments, the measurements of $P_{\rho \rt \rho}$
and $P_{\rho \rt \rho^{'}}$, can be used directly to calculate the
relaxation times $\tau_{\rho}$ and the transition matrix elements
$W_{\rho^{'}\rho}$ from $\rho$ to $\rho^{'}$:
\bea
 \sum_{\rho^{'}\neq \rho} W_{\rho^{'}\rho} &=& \tau_{\rho}^{-1} \nonumber \\
W_{\rho^{'}\rho} &=& \tau_{\rho}^{-1} {f_2 \over {1 - f_1}} 
\label{w_rates1}
\eea
The variation of the relaxation time scales with the tilt $\rho$ are
shown in Fig. \ref{tau_rho}$(a)$ for different values of 
$\mu$. The most striking feature of these time scales is that
$\tau_{\rho}$ increases exponentially with
$\rho$: $\tau_{\rho} \propto
\exp(\alpha ({\rho}^{2\gamma}))$ with $\gamma$ between $1/2$ and $1$.  
We will show below that the exponential increase
of $\tau_{\rho}$ arises from the jamming of non-zero winding number loops 
and is directly responsible for the Vogel-Fulcher 
divergence of the average time scale.

To explore the origin of the exponential rise of $\tau_{\rho}$ with
$\rho$, we investigated the elements of the transition matrix $W$
obtained from our measurements.  The rates for the transition
$W_{{\rho-1},\rho}$, and $W_{{\rho+1},\rho}$ are plotted in
Fig. \ref{tau_rho}$(b)$ and Fig. \ref{tau_rho}$(c)$, respectively for
different $\rho$'s as a function of $\mu$.
\figone{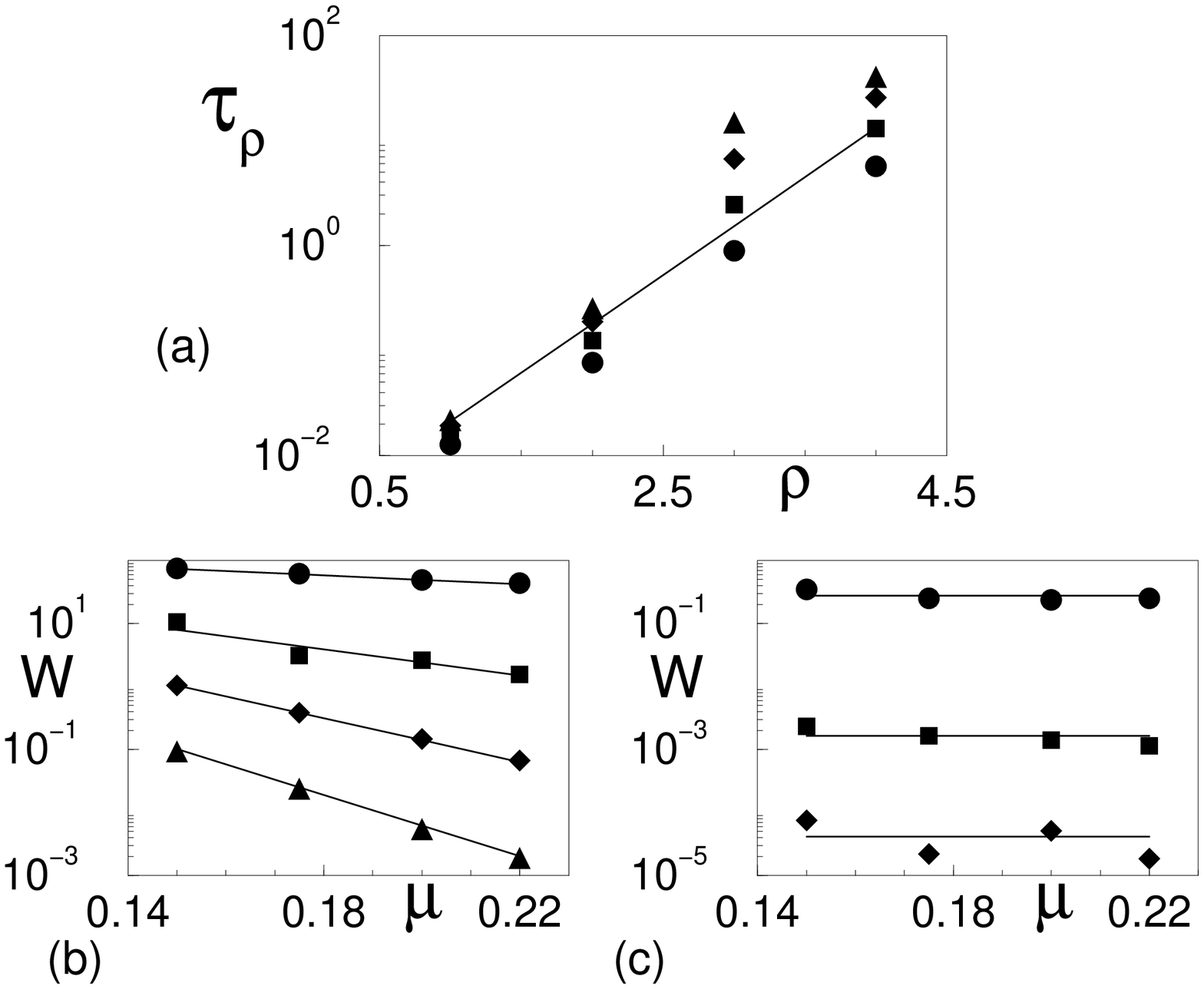}{We show (a) the exponential dependence of 
$\tau_{\rho}$ on $\rho$ for $\mu = 0.15$, $0.175$, $0.2$ and $0.22$.
The major contribution to the plateau timescale $\tau$ in $C(t)$ ({\it
cf}  Fig. \ref{autocorr}) comes from non-simple tilts.  For example, $\tau_{\rho}$ for $\brho = (4,2)$ is $\simeq 211$ and $660$ MCS,
for $\mu = 0.15$ and $0.175$, respectively.(b) $W_{\rho-1,\rho}$
versus $\mu$ at $\rho = 1$, $2$, $3$ and $4$ (from top to bottom), and
(c) $W_{\rho+1,\rho}$ versus $\mu$ at $\rho = 1$, $2$ and $3$ (from
top to bottom).}{tau_rho} 
%The excellent fits of these rates to an
%exponential of the form: 
Transition rates from higher to lower tilt are well fitted by   
$\exp{(-8 \mu (2\rho - 1))}$. This suggests these rates to be of the form:
\be
%W_{{\rho-1},{\rho}} = {{\Gamma_o \over {\Omega(\rho)}} e^{-(E_{\rho-1} - 
%E_{\rho}))}} 
W_{{\rho-1},{\rho}} = {{\Gamma_o} e^{-(E_{\rho-1} - E_{\rho}))}} 
\label{down}
\ee
The $\mu$ independent factor $\Gamma_o$ varies rather
weakly (about a factor of $2$ for the data in Fig. \ref{tau_rho}$(b)$)
with $\rho$, and thus the relaxation of $\rho$ to $(\rho
- 1)$ is dominated by the cost of going uphill in energy as expected.  

The unusual behavior of $\tau_{\rho}$ can be traced
to the transition rates $W_{{\rho+1},{\rho}}$ (Fig.~\ref{tau_rho}$(c)$) 
which can be written as:  
%These were found to
%have strong $\rho$ dependent blocking factors, $\Gamma(\rho)$, which
%suppressed the transitions taking the system downhill in energy:
\be
W_{{\rho + 1},\rho} = {\Gamma_o}{\Gamma (\rho)}.  
\label{up} 
\ee
The ``blocking factor'', $\Gamma(\rho)$, shows a negligible dependence on 
$\mu$ but a strong, exponential dependence on the tilt $\rho$.
The origin of $\Gamma(\rho)$ can be traced back to {\it jamming} caused 
by the system
spanning loops in a tilted state. These loops are meandering objects,
and their lengths scale as $L^{3/2}$ \cite{kondev,das} thus
squeezing in an extra loop in order to increase the tilt
is associated with straightening out 
the existing ones. Thus
the transition to higher tilt states are dramatically suppressed
with increasing $\rho$ and lead to an exponential increase of
$\tau_{\rho}$. Using Eqs.
\ref{w_rates1}, \ref{down} and \ref{up}, the time scales can be expressed as: 
\bea
\tau_{\rho} = {1 \over {W_{{\rho + 1},\rho} + W_{{\rho - 1},\rho}}} 
= {{1/\Gamma_o} \over {\Gamma(\rho) + e^{-({E_{\rho -1} - E_{\rho})}}}} 
\label{timescale1}
\eea
The $\rho$-dependent blocking factor in the denominator of
Eq. \ref{timescale1} cannot be scaled away and the dramatic
suppression of the the downhill transitions, reflected in
$\Gamma(\rho) \ll \exp(-(E_{\rho -1} - E_{\rho})) < 1$ 
leads to a $\tau_{\rho}$ being dominated by the exponential 
energy cost of decreasing the tilt\cite{foot1}.

\paragraph{Theory of Vogel-Fulcher divergence:}
We have established that the relaxation times $\tau_{\rho}$ grow
exponentially with $\rho$ because of barriers arising from the spatial
constraints imposed by the system-spanning loops. In this paragraph we discuss how the shape of $C(t)$ and the Vogel-Fulcher type divergence of $\tau$ can be related to the $\tau_{\rho}$.  We have good numerical evidence that 
the time dependence of $C(t)$ is well approximated  by the
form:
\be
C(t) \approx {{\sum_{\rho} e^{-{t/\tau_{\rho}}} {P_{eq}}(\rho)}
\over {\sum_{\rho} {P_{eq}}(\rho)}}
\label{corr_model}
\ee
This form of $C(t)$ is based on the approximation that the system is
perfectly correlated as long as it stays in a state with a given tilt
and decorrelates completely as it leaves this state.  The probability
of staying in a tilt $\rho$ for a time $t$ is measured by
$e^{-{t/\tau_{\rho}}}$ and ${P_{eq}}(\rho)$ measures the equilibrium
probability of the tilt $\rho$. The above model of $C(t)$ is
identical to that of the trap model of glasses\cite{bouchaud}.  Since
$\tau_{\rho}$ grows exponentially with the tilt $\sim
\exp(\alpha {\rho^{2\gamma}})$, the plateau timescale in $C(t)$ is dominated 
by the $\tau_{\rho}$ corresponding to the maximum accessible tilt,
$\rho_{max}$, as determined by the width of the distribution
$P_{eq}(\rho)$.  As the transition to the tilted state is approached,
the fluctuations of $\rho$ increase leading to a power-law divergence
of $\langle {\rho}^2 \rangle$ and $\rho_{max}$
and an exponential divergence (Vogel-Fulcher) of the plateau
timescale, $\tau$.

In summary, we find that a lattice model with frustration and
associated extended structures exhibits a trap dominated dynamics
arising due to jamming.  A phase transition involving these 
structures then naturally leads to an exponential divergence of timescales.

We thank J.P. Sethna, S. Kivelson, R. Stinchcombe, M. Barma, D. Dhar and
G. Tarjus for helpful discussions. This work was supported by the NSF
DMR-9815986 (DD, BC) and NSF DMR-9984471 (JK).

\end{multicols}


\begin{thebibliography}{99}
\bibitem{foot1}For all $\mu$'s studied we checked detailed balance, 
using the measured  $P^{eq}(\rho)$ and $W_{{\rho}^{'} \rho}$, and 
showed that the system was in equilibrium.
\end{thebibliography}
\end{document}